\documentclass[twocolumn,showpacs,preprintnumbers,amsmath,amssymb]{revtex4}
\usepackage{epsfig}
\usepackage{subfigure}
\usepackage{dcolumn}% Align table columns on decimal point
\usepackage{bm}% bold math
\usepackage[all, knot]{xy}
\usepackage{epstopdf}
\xyoption{arc}

\begin{document}

\title{Percolation in Networks with Voids and Bottlenecks\\ }% Force line breaks with \\

\author{Amir Haji-Akbari}
 \email{hajakbar@umich.edu}
\author{Robert M. Ziff}%
 \email{rziff@umich.edu}
\affiliation{%
Michigan Center for Theoretical Physics and Department of Chemical Engineering, University of Michigan, Ann
Arbor MI 48109-2136.
}%

\date{\today}% It is always \today, today,

\begin{abstract}
A general method is proposed for predicting the asymptotic percolation threshold of networks with bottlenecks, in the limit that the sub-net mesh size goes to zero. The validity of this method is tested for bond percolation on filled checkerboard and ``stack-of-triangle" lattices. Thresholds for the checkerboard lattices of different mesh sizes are estimated using the gradient percolation method, while for the triangular system they are found exactly using the triangle-triangle transformation. The values of the thresholds approach the asymptotic
values of $0.64222$ and $0.53993$ respectively as the mesh is made finer, consistent with a direct determination based upon the predicted critical corner-connection probability.
\end{abstract}

\maketitle

\section{\label{sec:Introduction}Introduction\protect\\ }

Percolation concerns the formation of long-range
connectivity in random systems~\cite{Stauffer}. It has a wide range of application in problems in physics and engineering, including such topics as conductivity and magnetism in random systems, fluid flow in porous media~\cite{Sukop2002}, epidemics and clusters in complex networks \cite{GoltsevDorogovtsevMendes08}, analysis of water structure \cite{BernabeiEtAl08},  and gelation in polymer systems \cite{YilmazGelirAlverogluUysal08}. To study this phenomenon, one typically models the network by a regular lattice made random by independently making sites or bonds occupied with probability $p$. At a critical threshold $p_c$, for a given lattice and percolation type (site, bond), percolation takes place. Finding that threshold exactly or numerically to high precision is essential to studying the percolation problem on a particular lattice, and has been the subject of numerous works over the years 
(recent works include Refs.\ \cite{Lee08,RiordanWalters07,Scullard06,ScullardZiff06,ZiffScullard06,ScullardZiff08,Parviainen07,QuintanillaZiff07,NeherMeckeWagner08,WiermanNaorCheng05,JohnerGrimaldiBalbergRyser08,KhamforoushShamsThovertAdler08,Ambrozic08,Kownacki08,FengDengBlote08,Wu06,MajewskiMalarz07,WagnerBalbergKlein06,TarasevichCherkasova07,HakobyanPapouliaGrigoriu07,BerhanSastry07}).

In this paper we investigate the percolation characteristics of networks with bottlenecks.  That is, we consider models in which we increase the number of internal bonds within a sub-net while keeping the number of contact points between sub-nets constant. We want to find how $p_c$ depends upon the mesh size in the sub-nets and in particular how it behaves as the mesh size goes to zero.  Studying such systems should give insight on the behavior of real systems with bottlenecks, like traffic networks, electric power transmission networks, and ecological systems. It is also interesting from a theoretical point of view because it interrelates the percolation characteristics of the sub-net and the entire network.

\begin{figure*}
\includegraphics[scale=0.4]{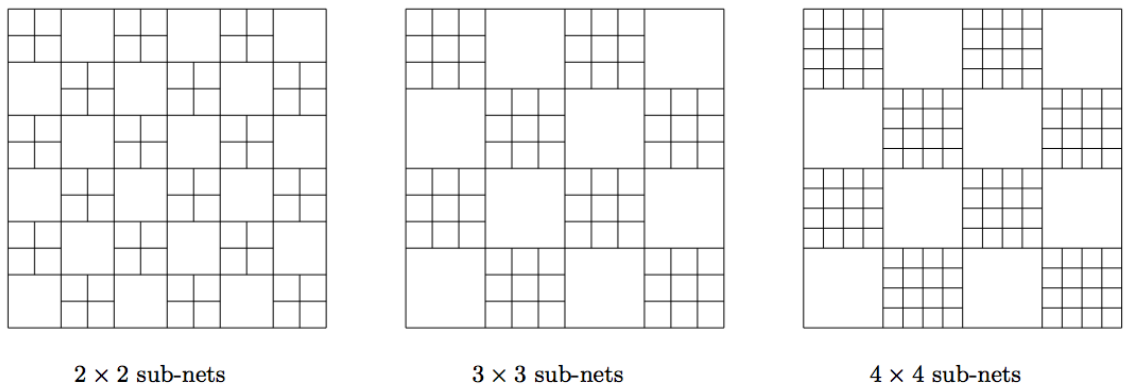}
\caption{\label{fig:Checkerboard_finite}Checkerboard lattices with sub-nets of finite sizes.}
\end{figure*}

\begin{figure*}
\includegraphics[scale=0.4]{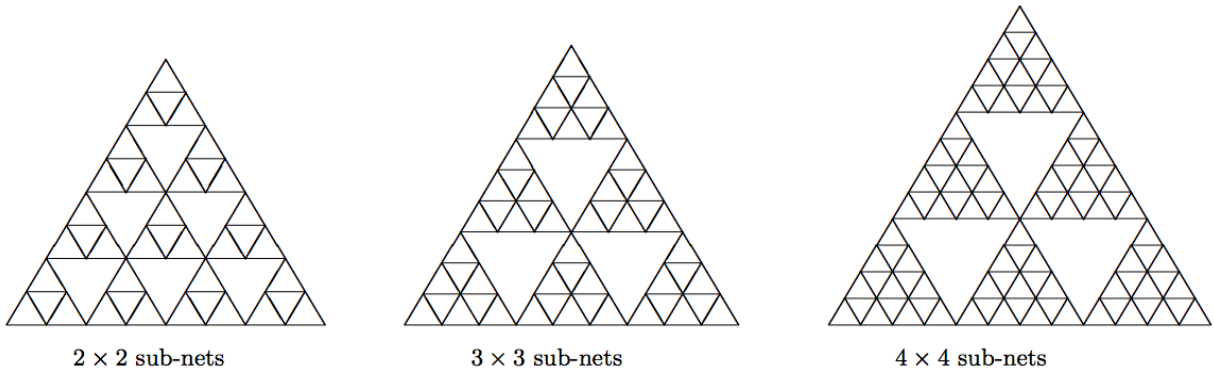}
 \caption{\label{fig:strg_finite}Stack-of-triangles lattice with sub-nets of finite sizes.}
\end{figure*}

An interesting class of such systems includes lattices with an ordered series of vacated areas within them. Examples include the filled checkerboard lattices (Fig.~\ref{fig:Checkerboard_finite}) and the ``stack-of-triangles" (Fig.~\ref{fig:strg_finite}).
The latter can be built by partitioning the
triangular lattice into triangular blocks of dimension $L$, and
alternately vacating those blocks. These internal blocks of length $L$ correspond to the sub-nets, which contact other sub-nets through the three contact points at their corners. The checkerboard lattice is the square-lattice analog of the stack-of-triangles lattice, where sub-nets are $L \times L$ square lattices which contact the other sub-nets via four contact points.  Note, for the 
stack-of-triangles sub-nets, we also use the $L \times L$ designation, here to indicate $L$ bonds on the base and the sides.

The problem of finding the bond percolation threshold can be solved exactly for the stack-of-triangles lattice because it fits into a class of self-dual arrangements of triangles, and the triangle-triangle transformation (a generalization of the star-triangle transformation) can be used to write down equations for its percolation threshold \cite{Ziff_CellDualCell,ChayesLei06}. This approach leads to an algebraic equation which can be solved using numerical root-finding methods.  However due to lack of self-duality in the filled checkerboard lattices, no exact solution can be obtained for their thresholds.

It is of interest and of practical importance to investigate the limiting behavior of systems with sub-nets of an infinite number of bonds, i.e., systems where the size of sub-nets is orders of magnitude larger than the size of a single bond in the system, or equivalently, where the mesh size of the lattice compared to the sub-net size becomes small.  Due to reduced connectivity, these systems will percolate at a \emph{higher} occupation probability than a similar regular lattice. The limiting percolation threshold for infinite sub-nets is counter-intuitively non-unity, and is argued to be governed by the connectedness of contact points to the infinite percolating clusters within sub-nets.  This argument leads to a simple criterion linking the threshold to the probability that the corners connect to the giant cluster in the center of the sub-net.

In this work, the limiting threshold value is computed for bond percolation on the stack-of-triangles and filled checkerboard lattices using this new criterion. Percolation thresholds are also found for a series of lattices of finite sub-net sizes. For the stack-of-triangles lattices, most percolation thresholds are evaluated analytically using the triangle-triangle transformation method, while for filled checkerboard lattices, the gradient percolation method~\cite{Ziff_Sapoval} is used. The limiting values of $0.53993$ and $0.64222$ are found for percolation thresholds of stack-of-triangles and checkerboard lattices respectively, which are both in good agreement with the values extrapolated for the corresponding lattices of finite sub-net sizes.

We note that there are some similarities between this work and studies done on the fractal Sierpi\'nski gaskets (triangular)~\cite{YuYao1988} and carpets (square), but in the case of the Sierpi\'nski models, the sub-nets are repeated in a hierarchical fashion while here they are not.  For the Sierpi\'nski gasket, which is effectively all corners, the percolation threshold is known to be 1 \cite{GefenAharonyShapirMandelbrot84}. For Sierpi\'nski gaskets of a finite
number of generations, the formulae for the corner connectivities can be found exactly through recursion
\cite{TaitelbaumHavlinGrassbergerMoenig90}, while here they cannot.  Recently another hierarchical model with bottlenecks, the so-called Apollonian networks, which are related to duals of Sierpinski networks, has also been introduced \cite{AutoMoreiraHerrmannAndrade08}.  In this model, the percolation threshold goes to zero as the system size goes to infinity.

\section{\label{sec:theory}Theory\protect\\}
Let $p$ be the probability that a bond in the system is occupied. Consider a network with sub-nets of infinitely fine mesh, each individually percolating (in the sense of forming ``infinite" clusters but not necessarily connecting the corners) at $p_{c,s}$, and denote the overall bond percolation threshold of the entire network to be $p_{c,n}$. It is obvious that $p_{c,s}<p_{c,n}$, due to reduced connectivity in the entire network compared to connectivity in individual sub-nets. For $p_{c,s} < p < p_{c,n}$, an infinite cluster will form within each sub-net with probability $1$. However, the entire network will not percolate, because a sufficient number of connections has not yet been established between the contact points at the corners and central infinite clusters.

Now we construct an auxiliary lattice by connecting the contact points to the center of each subnet, which represents the central infinite cluster contracted into a single site. The occupation probability of a bond on this auxiliary lattice is the probability that the contact point is connected
to the central infinite cluster of the sub-net. Percolation of this auxiliary lattice is equivalent to the percolation of the entire network. That is,  if this auxiliary lattice percolates at a threshold $p_{c,a}$, the percolation threshold of the entire network will be determined by:
\begin{eqnarray}
\label{eq:bottleneck_general}
P_{\infty,{\rm corner}}(p_{c,n})=p_{c,a}
\end{eqnarray}
where $P_{\infty,{\rm corner}}(p)$ gives the probability that the corner of the sub-net is connected to the central infinite cluster given that the single occupation probability is $p$.

In general no analytical expression exists for $P_{\infty,{\rm corner}}(p)$, even for simple lattices such as the triangular and square lattices, and $P_{\infty,{\rm corner}}(p)$ must be evaluated by simulation.

\begin{figure}
\includegraphics[scale=0.7]{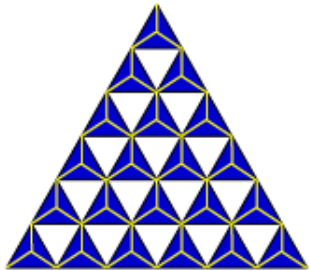}
\caption{\label{fig:stack_of_triangles} (Color online.)  Stack-of-triangles lattice and its
auxiliary lattice.  The filled blue (dark) triangles represent the sub-net, and the yellow honeycomb
lattice represents the effective auxiliary lattice.}
\end{figure}

\subsection{\label{sec:theory:stack_of_triangles}Stack-of-Triangles Lattice}

Fig.~\ref{fig:stack_of_triangles} shows a limiting stack-of-triangles lattice where each shaded triangle represents a sub-net of infinitely many bonds. The contact points are the corners of the triangular sub-nets. As shown in Fig.~\ref{fig:stack_of_triangles}, the auxiliary lattice of the stack-of-triangles lattice is the honeycomb lattice, which percolates at ${p_{c,a}=1-2\sin\left(\pi/18\right)\approx0.652704}$~\cite{SykesEssam1964}. Thus the asymptotic percolation threshold $p_{c,n}$ of the stack-of-triangles will be determined by:
\begin{eqnarray}
\label{eq:bottleneck_str}
P_{\infty,{\rm corner}}(p_{c,n})=1-2\sin{\frac{\pi}{18}} \ .
\end{eqnarray}
Because the stack-of-triangles lattice is made up of triangular cells in a self-dual arrangement, its percolation threshold can be found exactly using the triangle-triangle transformation \cite{Ziff_CellDualCell, ChayesLei06}. Denoting the corners of a single triangular sub-net with $A$, $B$ and $C$, the percolation threshold of the entire lattice is determined
by the solution of the following equation:
\begin{eqnarray}
\label{eq:dual}
P(ABC)=P(\overline{ABC})
\end{eqnarray}
where ${P(ABC)}$ is the probability that $A$, $B$ and $C$ are all connected, and ${P(\overline{ABC})}$ is the probability that none of them are connected. Eq.~(\ref{eq:dual}) gives rise to an algebraic equation which can be solved for the exact percolation threshold of the lattices of different sub-net sizes.

\subsection{\label{sec:theory:checkerboard}Filled Checkerboard Lattice}

\begin{figure}
\includegraphics[scale=0.5]{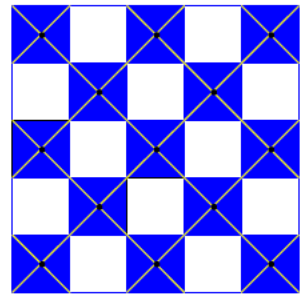}
\caption{\label{fig:checkerboad_aux}(Color online.) Auxiliary lattice of the checkerboard
lattice.  The blue (dark) colored areas represent the subnets, and the double-bond square
lattice (diagonals) represents the auxiliary lattice.}
\end{figure}

Unlike the stack-of-triangles lattice, there is no exact solution for percolation threshold of the checkerboard lattice for finite sub-nets because no duality argument can be made for such lattices. However once again an auxiliary lattice approach can be used to find a criterion for the asymptotic value of percolation threshold. Fig.~\ref{fig:checkerboad_aux} depicts the corresponding auxiliary lattice for a checkerboard lattice, which is simply the square lattice with double bonds in series.  This lattice percolates at ${p_{c,a}= 1/\sqrt{2} \approx{0.707107}}$. Thus for the infinite sub-net ${p_{c,n}}$ will be determined by:
\begin{eqnarray}
\label{eq:bottleneck_checkerboard}
P_{\infty,{\rm corner}}(p_{c,n})= \frac{1}{\sqrt{2}}
\end{eqnarray}
It is interesting to note that there exists another regular lattice --- the ``martini" lattice --- for which the bond threshold is also exactly $1/\sqrt{2}$ \cite{ZiffScullard06}.  However, that lattice does not appear to relate to a network construction as the double-square lattice does.
%\begin{figure}
%\includegraphics[scale=0.6]{Checkerboard.pdf}
%\caption{\label{fig:checkerboard}Checkerboard lattice with sub-nets of
%$4$-by-$4$ square lattices.}
%\end{figure}

\section{\label{sec:methods}Methods}

\subsection {\label{sec:methods:Pc_finite}Percolation Threshold of Systems of finite-sized sub-nets}

For the checkerboard lattice, we estimate the bond percolation thresholds using the gradient percolation method ~\cite{Ziff_Sapoval}. In this method, a gradient of occupation probability is applied to the lattice, such that bonds are occupied according to the local probability determined by this gradient. A self-avoiding hull-generating walk is then made on the lattice according to the rule that an occupied bond will reflect the walk while a vacant bond will be traversed by the walk. For a finite gradient, this walk can be continued infinitely by replicating the original lattice in the
direction perpendicular to the gradient using periodic boundary conditions. Such a walk will map out the boundary between the
percolating and non-percolating regions, and the average value of occupation probability during the walk will be a measure of the percolation threshold. Because all bonds are occupied or vacated independent of each other, this average probability can be estimated as \cite{RossoGouyetSapoval1986}:
\begin{eqnarray}
\label{eq:Pc_gradient}
p_c=\frac{N_{occ}}{N_{occ}+N_{vac}}
\end{eqnarray}
It is particularly straightforward to implement this algorithm to bond percolation on a square lattice, and the checkerboard lattice can be simulated by making some of the square-lattice bonds permanently vacant.  Walks are carried out in a horizontal-vertical direction and the original lattice is rotated $45^{\circ}$.

\begin{figure}[htbp] %  figure placement: here, top, bottom, or page
   \centering
   \includegraphics[width=2.5 in]{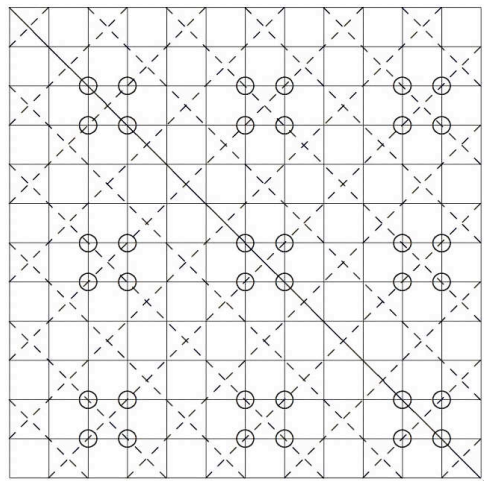} 
   \caption{\label{fig:chkrbrd_2b2}Representation of checkerboard lattices for simulation with gradient method.  The original bond lattice is represented by 
   dashed diagonal lines, while lattice on which the walk goes is vertical and horizontal.  Open circles mark bonds that are permanently vacant.}

\end{figure}

\begin{figure}[htbp] %  figure placement: here, top, bottom, or page
   \centering
   \includegraphics[width=2.5 in]{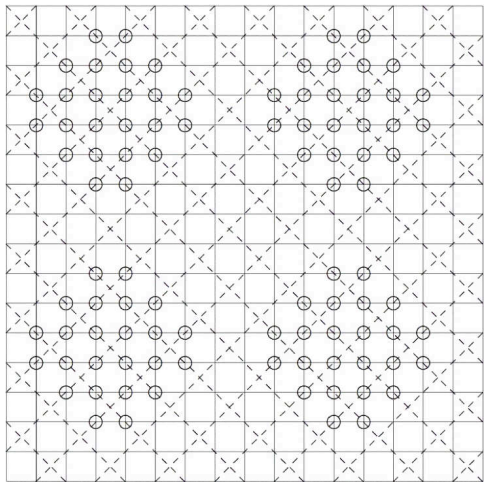} 
 \caption{\label{fig:chkrbrd_4b4}Checkerboard lattice with $4\times4$ vacancies, with description the same as in Fig.\ \ref{fig:chkrbrd_2b2}.}
\end{figure}

We applied this approach to checkerboard lattices
of different block sizes. Fig.~\ref{fig:chkrbrd_2b2} and
Fig.~\ref{fig:chkrbrd_4b4} show the corresponding setups for lattices with $2\times2$ and $4\times4$ vacancies, where the lattice bonds are represented as dashed diagonal lines and solid horizontal and vertical lines show where the walk goes. Circles indicate the centers of permanently vacant bonds. It should be emphasized that permanently vacated bonds are not counted in Eq.\ (\ref{eq:Pc_gradient}) even if they are visited by the walk.

The percolation threshold of stack-of-triangles lattices of finite sub-net size were calculated using Eq.\ (\ref{eq:dual}). If the occupation probability is $p$ and $q = 1 - p$,  one can express $P(ABC)$ and $P(\overline{ABC})$ as:
\begin{eqnarray}
P(ABC)&=&\sum_{i=0}^{3n(n+1)/2} \phi(n,i)p^iq^{3n(n+1)/2-i}
\label{eq:phi_n_i}\\
P(\overline{ABC})&=&\sum_{i=0}^{3n(n+1)/2}\psi(n,i)p^i q^{3n(n+1)/2-i}
\label{eq:psi_n_i}
\end{eqnarray}
where $n$ denotes the number of bonds per side of the sub-net, $\phi(n,i)$ denotes the number of configurations of an $n\times n$ triangular block with precisely $i$ occupied bonds where the $A$, $B$ and $C$ are connected to each other and $\psi(n,i)$ denotes the number of configurations where none of these points are connected. There appears to be no closed-form combinatorial expression for $\phi(n,i)$ and $\psi(n,i)$, and we determined them by exhaustive search of all possible configurations.

%\begin{figure*}
%\includegraphics{Checkerboard_2b2.pdf}% Here is how to import %\caption{\label{fig:chkrbrd_2b2}Checkerboard lattice with $2\times2$ vacancies rotated by ${45^{\circ}}$. Crosses correspond to permanently-vacated bonds reflecting the holes in the system. The original lattice is represented with diagonal bonds.}
%\end{figure*}

%\begin{figure*}
%\includegraphics{Checkerboard_4b4.pdf}% Here is how to import EPS art
%\caption{\label{fig:chkrbrd_4b4}Checkerboard lattice with $4\times4$ vacancies. Crosses correspond to permanently vacated bonds. The original lattice is represented with diagonal bonds.}
%\end{figure*}

\subsection {\label{sec:methods:estimate_Pinf}Estimation of ${P_{\infty,{\rm corner}}}$}
As mentioned in Section \ref{sec:theory}, the asymptotic value of percolation threshold ${p_{c,n}}$  can be calculated using
Eq.\ (\ref{eq:bottleneck_checkerboard}). However there is no analytical expression for ${P_{\infty,{\rm corner}}(p)}$, hence it must be characterized by simulation. In order to do that, the size distribution of clusters connected to the corner must be found for different values of $p>p_{c,s}$. Cluster sizes are defined in terms of the number of sites in the cluster.

In order to isolate the cluster connected to the corner, a first-in-first-out Leath or growth algorithm is used starting from the corner. In FIFO algorithm, the neighbors of every unvisited site are investigated before going to neighbors of the neighbors, so that clusters grow in a circular front. Compared to last-in-first-out algorithm used in recursive programming, this algorithm performs better for ${p{\ge}p_{c,s}}$ because it explores the space in a more compact way.

\begin{figure*}[htp]
\centering
%\subfigure[$p=0.40$]{\label{fig:clst_040}}
\subfigure[]{\label{fig:clst_040}}
\includegraphics[scale=0.25]{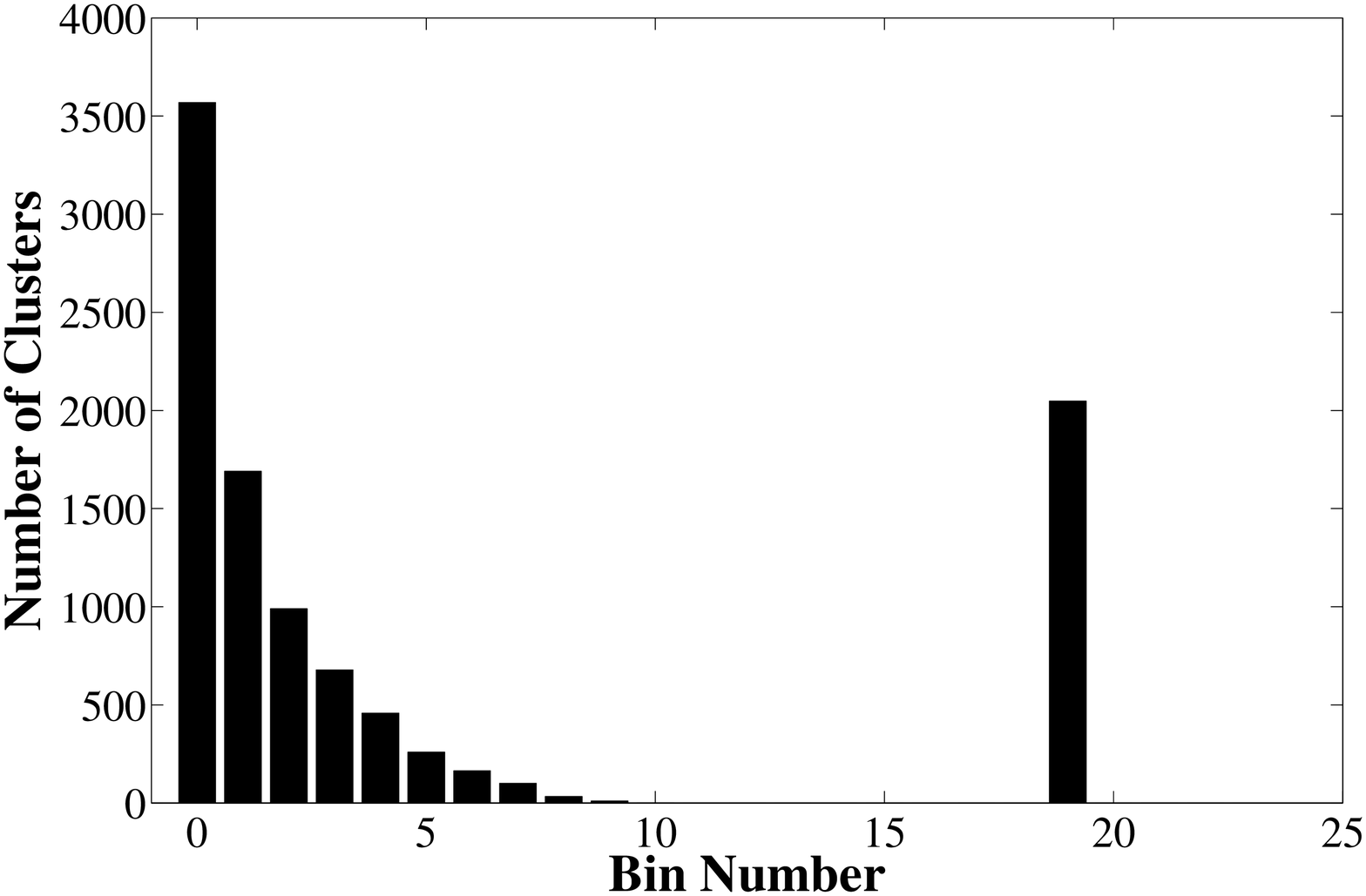}
%\hspace{1.2in} \subfigure[$p=0.55$]{\label{fig:clst_055}} \hspace{-.1in} \vspace{.1in}
\hspace{0.5in} \subfigure[]{\label{fig:clst_055}} \hspace{-.1in} \vspace{.1in}
\includegraphics[scale=0.25]{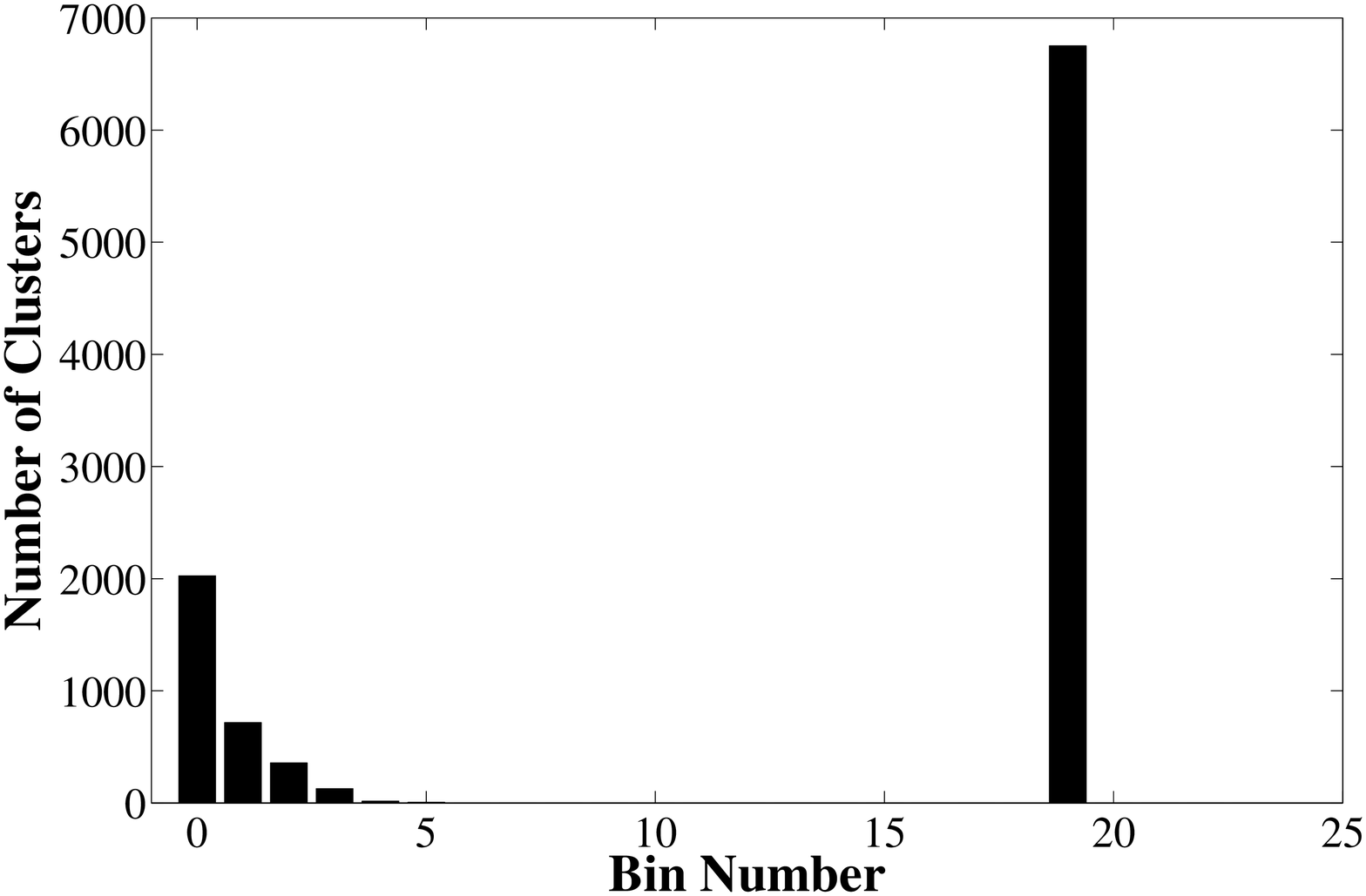}
\caption{\label{fig:cluster_size}Cluster size distribution for clusters originating from the corner of triangular lattice at
(a): $p=0.40$ and (b): $p=0.55$, after $10^4$ independent runs.}
\end{figure*}

At each run, the size of the cluster connected to the corner is evaluated using the FIFO growth algorithm. In order to get better statistics, clusters with sizes between ${2^i}$ and ${2^{i+1}-1}$ are counted to be in the $i-$th bin. Because simulations are always run on a finite system, there is an ambiguity on how to define the infinite cluster. However, when ${p{\ge}p_{c,s}}$, the infinite cluster occupies almost the entire lattice, and the finite-size clusters are quite small on average. This effect becomes more and more profound as $p$ increases, and the expected number of clusters in a specific bin becomes smaller and smaller. Consequently, larger bins will effectively contain no clusters, except the bin corresponding to cluster sizes comparable to the size of the entire system. Thus there is no need to set a cutoff value for defining an infinite cluster. Fig.~\ref{fig:cluster_size} depicts the size distribution of clusters connected to the corner obtained for $1024 \times 1024$
  triangular lattice at (a): $p=0.40$ and (b): $p=0.55$ after ${10^4}$ independent runs. As it is observed, there is a clear gap between bins corresponding to small clusters and the bin corresponding to the spanning infinite cluster even for small values of $p$, which clearly demonstrates that the largest nonempty bin corresponds to infinite percolating clusters connected to the corner.  The fraction of such clusters connected to the corner is an estimate of $P_{\infty,{\rm corner}}(p)$.
  
In the simulations, we used the four-offset shift-register random-number generator R(471,1586,6988,9689)  described in Ref.\ \cite{Ziff98}.

\section{\label{sec:results}Results and Discussion}

\subsection{\label{sec:results:gradient}Gradient Percolation Data}
\begin{figure*}[htp]
\centering
\subfigure[]{\label{fig:pc_2b2}}
\includegraphics[scale=0.25]{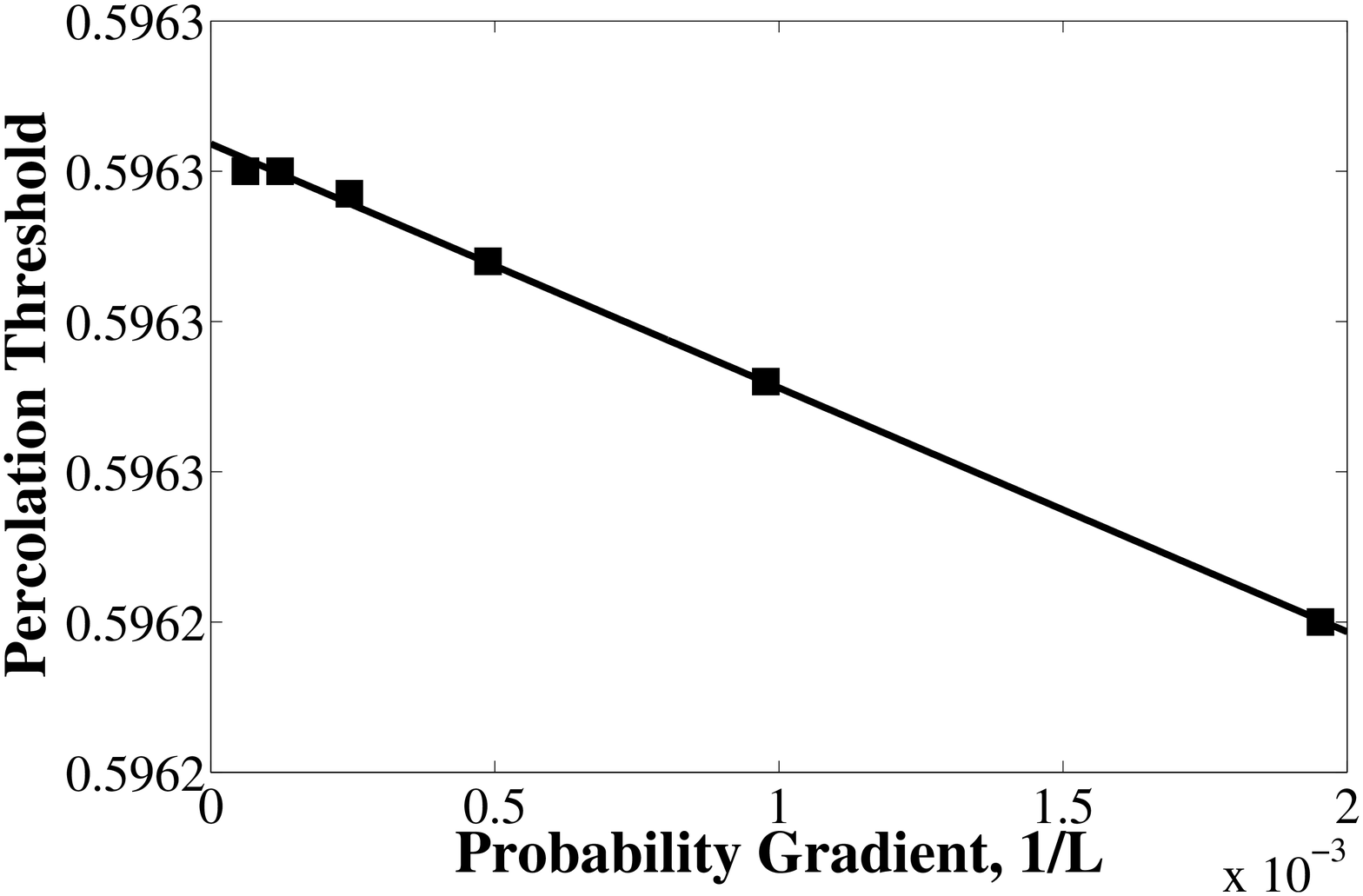}
\hspace{0in} \subfigure[]{\label{fig:pc_4b4}} \hspace{-.1
in} \vspace{.1in}
\includegraphics[scale=0.25]{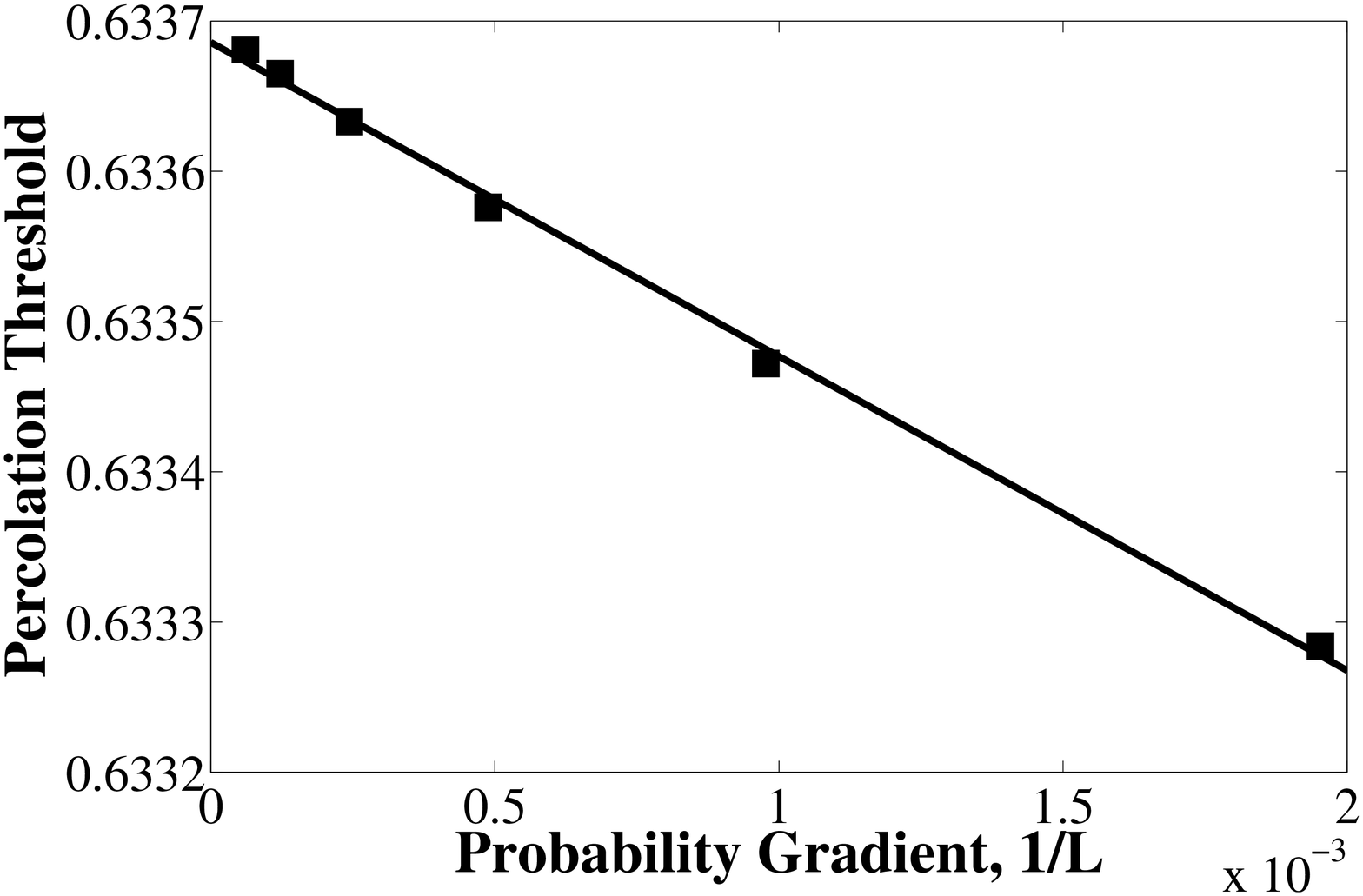}
\subfigure[]{\label{fig:pc_8b8}}\hspace{-.1in}\vspace{.1in}
\includegraphics[scale=0.25]{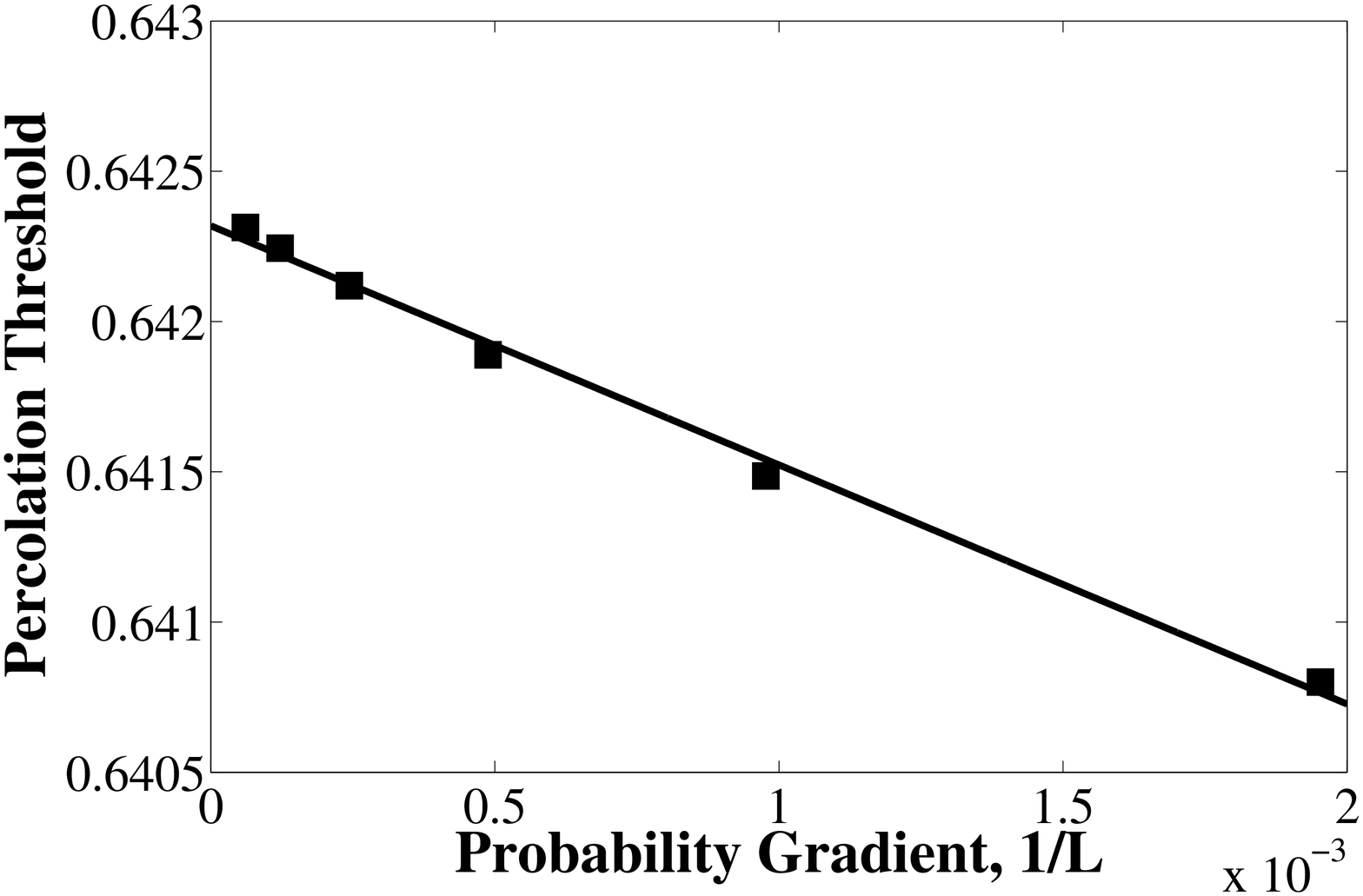}
\caption{\label{fig:pc_checkerboard}Gradient percolation data for checkerboard lattices with: (a) $2\times2$, (b) $4\times4$, and (c) $8\times8$ sub-net sizes. Squares correspond to the bond percolation thresholds estimated for different values of probability gradients; error bars are smaller than the symbol size.}
\end{figure*}

The gradient percolation method was used to estimate the bond percolation threshold of checkerboard lattices of five different sub-net sizes, i.e., $2\times2$, $4\times4$, $8\times8$, $16\times16$ and $32\times32$. For each lattice, six values of the gradient were used, and simulations were run for $10^{10}$ to $10^{12}$ steps for each gradient value in order to assure that the estimated percolation thresholds are accurate to at least five significant digits. The gradient was applied at an angle of $45^{\circ}$ relative to the original lattice. Figures~\ref{fig:pc_2b2}-~\ref{fig:pc_8b8} depict typical simulation results. Measured percolation thresholds for finite gradients were extrapolated to estimate the percolation threshold as $L\rightarrow\infty$. Our simulations show that $p_c$ fits fairly linearly when plotted against $1/L$. Table~\ref{table:pc_checkerboard} gives these estimated percolation thresholds.

\begin{table}
\caption{\label{table:pc_checkerboard} Percolation threshold for checkerboard lattices of different sub-net sizes:
 $^a$Exact result, $^{b}$from gradient percolation simulations, $^c$from corner simulations using Eq.\ (\ref{eq:bottleneck_checkerboard}).}
\begin{tabular}{ccccccc}
    \hline
    \hline
    Sub-net size & & & & & & Estimated $p_{c,n}$ \\
    \hline
    $1\times1$ & & & & & & $0.5^{a}$ \\
    $2\times2$ & & & & & & $0.596303\pm0.000001^{b}$ \\
    $4\times4$ & & & & & & $0.633685\pm0.000009^{b}$ \\
    $8\times8$ & & & & & & $0.642318\pm0.000005^{b}$ \\
    $16\times16$& & & & & & $0.64237\pm0.00001^{b}$ \\
    $32\times32$& & & & & & $0.64219\pm0.00002^{b}$ \\
    $\vdots$ & & & & & & $\vdots$\\
    $\infty$ & & & & & & $0.642216\pm0.00001^{c}$ \\
    \hline
\end{tabular}
\end{table}

\subsection{\label{section:results:dual}Percolation Threshold of The Stack-of-triangles Lattice}

As mentioned in Sections~\ref{sec:theory:stack_of_triangles} and~\ref{sec:methods:Pc_finite}, the percolation threshold
of stack-of-triangles lattice can be determined by Eq.\ (\ref{eq:dual}). Table~\ref{table:polynomial} summarizes the corresponding polynomial expressions and their relevant roots for lattices having $1$, $2$, $3$ and $4$ triangles per edge.  These polynomials give the $\phi(n,i)$ and $\psi(n,i)$ in Eqs.\ (\ref{eq:phi_n_i}) and (\ref{eq:psi_n_i}) for $n=1,2,3,4$ and $i=0,1,\ldots,3n(n+1)/2$.   We show $p_0 = P(\overline{ABC})$, $p_2 = P(AB\overline{C})$ (the probability that a given pair of vertices are connected together and not connected to the third vertex), and $p_3 = P(ABC)$.   These quantities satisfy $p_0 + 3 p_2 + p_3 = 1$.  Then we use Eq.\ (\ref{eq:dual}) to solve for $p_{c,n}$ numerically.

We also show in Table \ref{table:polynomial}  the values of $p_0$, $p_2$ and $p_3$ evaluated at the $p_{c,n}$.  Interestingly, as $n$ increases, $p_0$ at first increases somewhat but then tends back to its original value at $n= 1$, reflecting the fact that the connectivity of the infinitely fine mesh triangle is identical to that of the critical honeycomb lattice, which is identical to the connectivity of the simple triangular lattice according to the usual star-triangle arguments.  

It is not possible to perform this exact enumeration for larger sub-nets, so we used gradient percolation method to evaluate $p_c$ for $5\times5$.  (To create the triangular bond system on a square bond lattice, alternating horizontal bonds are made permanently occupied.)  The final threshold results are summarized in Table \ref{table:pc_stack_of_triangles}.

\begin{table*}
\caption{\label{table:polynomial}Exact enumeration polynomials and $p_c = p_{c,n}$ for the stack-of-triangles lattices }
\begin{ruledtabular}
\begin{tabular}{lll}
Sub-net & $1\times1$ (simple triangular lattice) & $2\times2$ (3 ``up" triangles or 9 bonds per sub-net)\\
\hline
${p_3=P(ABC)}$ & ${p^3+3p^2q}$ & ${9p^4q^5+57p^5q^4+63p^6q^3+33p^7q^2+9p^8q+p^9}$\\
$p_2 =P\left(AB\overline{C}\right)$ & ${pq^2}$ & ${p^2q^7+10p^3q^6+32p^4q^5+22p^5q^4+7p^6q^3+p^7q^2}$ \\
${p_0=P(\overline{ABC})}$ & ${(p+q)^3-p_3-3p_2}$ & ${(p+q)^9-p_3-3p_2}$\\
$p_c$ &$0.34729635533$ & $0.47162878827$ \\
${p_0(p_c)=p_3(p_c)}$ & $0.27806614328$ & $0.28488908000$ \\
${p_2(p_c)}$ & $0.14795590448$ & $0.14340728000$ \\
\end{tabular}
\begin{tabular}{ll}
Sub-net & $3 \times 3$ (6 ``up" triangles or $18$ bonds per sub-net)\\
\hline ${p_3=P(ABC)}$ &
${29p^6q^{12}+468p^7q^{11}+3015p^8q^{10}+9648p^9q^9+16119p^{10}q^8+17076p^{11}q^7+12638p^{12}q^6}$
\\
{} & ${+6810p^{13}q^5+2694p^{14}q^4+768p^{15}q^3+150p^{16}q^2+18p^{17}q+p^{18}}$ \\
  % & \\
$p_2 =P\left(AB\overline{C}\right)$ & ${p^3q^{15}+21p^4q^{14}+202p^5q^{13}+1125p^6q^{12}+3840p^7q^{11}+7956p^8q^{10}+9697p^9q^9}$ \\
{} & ${+7821p^{10}q^8+4484p^{11}q^7+1879p^{12}q^6+572p^{13}q^5+121p^{14}q^4+16p^{15}q^3+p^{16}q^2}$ \\
% & \\
${p_0=P(\overline{ABC})}$ & ${(p+q)^{18}-p_3-3p_2}$\\
$p_c$ &$0.50907779266$  \\
${p_0(p_c)=p_3(p_c)}$ & $0.28322276251$  \\
${p_2(p_c)}$ & $0.14451815833$ \\
\end{tabular}
\begin{tabular}{ll}
Sub-net & $4 \times 4$ ($10$ ``up" triangles or $30$ bonds per sub-net) \\
\hline ${p_3=P(ABC)}$ & ${99p^8q^{22}+2900p^9q^{21}+38535p^{10}q^{20}+305436p^{11}q^{19}+1598501p^{12}q^{18}}$  \\
{} & ${+5790150p^{13}q^{17}+14901222p^{14}q^{16}+27985060p^{15}q^{15}+39969432p^{16}q^{14}}$ \\
{} & ${+45060150p^{17}q^{13}+41218818p^{18}q^{12}+31162896p^{19}q^{11}+19685874p^{20}q^{10}}$ \\
{} & ${+10440740p^{21}q^9+4647369p^{22}q^8+1727208p^{23}q^7+530552p^{24}q^6+132528p^{25}q^5}$ \\
{} & ${+26265p^{26}q^4+3976p^{27}q^3+432p^{28}q^2+30p^{29}q+p^{30}}$ \\
%  & \\
$p_2 =P\left(AB\overline{C}\right)$ & ${p^4q^{26}+36p^5q^{25}+613p^6q^{24}+6533p^7q^{23}+48643p^8q^{22}+267261p^9q^{21}}$ \\
{} & ${+1114020p^{10}q^{20}+3563824p^{11}q^{19}+8766414p^{12}q^{18}+16564475p^{13}q^{17}}$ \\
{} & ${+24187447p^{14}q^{16}+27879685p^{15}q^{15}+25987202p^{16}q^{14}+19980934p^{17}q^{13}}$ \\
{} & ${+12843832p^{18}q^{12}+6950714p^{19}q^{11}+3170022p^{20}q^{10}+1212944p^{21}q^9}$ \\
{} & ${+385509p^{22}q^8+100140p^{23}q^7+20744p^{24}q^6+3300p^{25}q^5+379p^{26}q^4+28p^{27}q^3+p^{28}q^2}$ \\
%  &  \\
${p_0=P(\overline{ABC})}$ & ${(p+q)^{30}-p_3-3p_2}$\\
$p_c$ &$0.52436482243 $  \\
${p_0(p_c)=p_3(p_c)}$ & $0.28153957013$  \\
${p_2(p_c)}$ & $0.14564028658$ \\
\end{tabular}
\end{ruledtabular}
\end{table*}

\begin{table}
\caption{\label{table:pc_stack_of_triangles} Percolation threshold for stack-of-triangles lattices of different sub-net sizes: $^a$From Eq.\ (\ref{eq:dual}) using exact expressions for $p_0$ and $p_3$ from Table \ref{table:polynomial}, $^b$from gradient simulation, $^c$Corner simulation using Eq.\ (\ref{eq:bottleneck_str}).}
\begin{tabular}{ccccccc}
    \hline
    \hline
    Sub-net size & & & & & & Estimated $p_c$ \\
    \hline
    $1\times1$ & & & & & & $0.347296355^{a}$ \\
    $2\times2$ & & & & & & $0.471628788^{a}$ \\
    $3\times3$ & & & & & & $0.509077793^{a}$ \\
    $4\times4$ & & & & & & $0.524364822^{a}$ \\
    $5\times5$ & & & & & & $0.5315976\pm0.000001^{b}$\\
    $\vdots$ & & & & & & $\vdots$\\
    $\infty$ & & & & & & $0.53993\pm0.00001^c$ \\
    \hline
\end{tabular}
\end{table}

\subsection{\label{sec:results:estimate_Pinf}Estimation of ${P_{\infty,{\rm corner}}(p)}$}

\subsubsection{\label{sec:results:estimate_Pinf:checkerb}Square Lattice}

\begin{figure}
\includegraphics[scale=0.25]{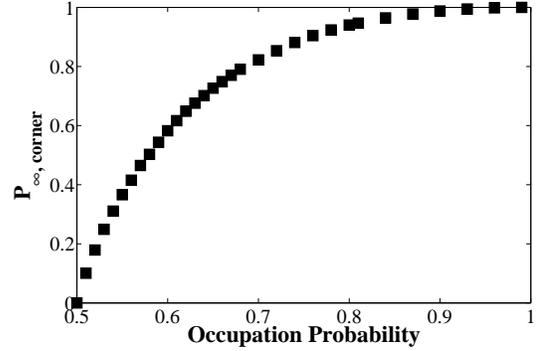}
\caption{\label{fig:p_inf_chkrbrd}${P_{\infty,{\rm corner}}(p)}$ for the square lattice.}
\end{figure}

The cluster growth algorithm was used to estimate ${P_{\infty,{\rm corner}}(p)}$ for different values of $p$. Simulations were run on a $2048\times2048$ square lattice. For each value of $p>1/2$, $10^5$ independent runs were performed and $P_{\infty,{\rm corner}}$ was estimated by considering the fraction of clusters falling into the largest
nonempty bin as described in Section~\ref{sec:methods:estimate_Pinf}. Fig.~\ref{fig:p_inf_chkrbrd} demonstrates the resulting curve for the square lattice. In order to solve Eq.\ (\ref{eq:bottleneck_checkerboard}), a cubic spline with natural boundary conditions was used for interpolation, and an initial estimate of ${p_{c,n}}$ was obtained to be $0.6432$. The standard deviation of $P_{\infty,{\rm corner}}(p)$ scales as $O(1/\sqrt{N})$ where $N$ is the number of independent simulation used for its estimation, so that  $N=10^5$ will give us an accuracy in $P_{\infty,{\rm corner}}(p)$ of about two significant figures.

In order to increase the accuracy in our estimate, further simulations were performed at the vicinity of $p=0.6432$ for $N=10^{10}$ trials with a lower cut-off size, and $p_{c,n}$ was found to be $0.642216\pm0.00001$. This number is in good agreement with percolation thresholds given in Table ~\ref{table:pc_checkerboard}. Note that $p_{c,n}$ for the $16\times16$ sub-net checkerboard lattice actually overshoots the value 0.642216 for the infinite sub-net and then drops to the final value.  This non-monotonic behavior is surprising at first and presumably is due to some interplay between the various corner connection probabilities that occurs for finite system.

At the threshold $p_{c,n} = 0.642216$, we found that the number of corner clusters containing $s$ sites for large $s$ behaves in the expected way for supercritical clusters \cite{Stauffer}  $n_s \sim a \exp(-b s^{1/2})$ with  $\ln a = -7.0429$ and $b = -0.8177$.

\subsubsection{\label{sec:results:estimate_Pinf:stotr}Triangular lattice}

\begin{figure}
\includegraphics[scale=0.25]{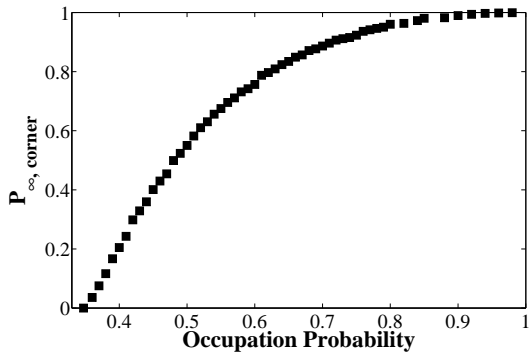}
\caption{\label{fig:p_inf_strg}${P_{\infty,{\rm corner}}(p)}$ for the triangular lattice.}
\end{figure}

The cluster growth algorithm was applied to find the size distribution of clusters connected to the corner of a $1024\times1024$ triangular lattice. For each value of $p$, $10^4$ independent runs were performed and $P_{\infty,{\rm corner}}(p)$ was evaluated. Fig.~\ref{fig:p_inf_strg} depicts the results. The root of Eq.\ (\ref{eq:bottleneck_str}) was determined by cubic spline to be around $0.539$.   Further simulations were performed around this value with $N=10^{10}$ runs for each $p$, yielding $p_{c,n}=0.539933\pm0.00001$. This value is also in good agreement with values given in Table ~\ref{table:polynomial} and shows fast convergence as sub-net size increases.

\section{Discussion}
\label{sec:conclusion}

We have shown that the percolation threshold of checkerboard and stack-of-triangle systems approach values less than 1 as the mesh spacing in the sub-nets goes to zero.   In that limit, the threshold can be found by finding the value of $p$ such that the probability a corner vertex is connected to the infinite cluster $P_{\infty,{\rm corner}}$ equals $1/\sqrt{2}$ and $1 - 2 \sin (\pi/18)$, respectively, based upon the equivalence with the double-bond square and bond honeycomb lattices.  The main results of our analysis and simulations are summarized in Tables \ref{table:pc_checkerboard} and \ref{table:pc_stack_of_triangles}.

\begin{figure}
\includegraphics[width = 2.5 in]{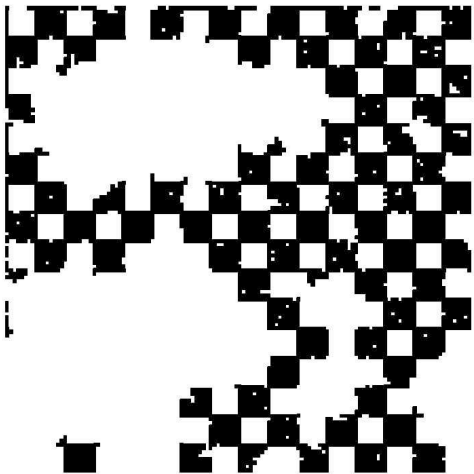}
\caption{\label{Pict8x8squaresBW}  A particular realization of the $8\times8$ checkerboard lattice at its critical point $p_{c,n} = 0.642318$ (as given in Table \ref{table:pc_checkerboard}), showing in black the sites wetted by the largest cluster.  The total
lattice is $128\times128$ sites, with periodic b.c.}
\end{figure}

For the case of the checkerboard, we notice a rather interesting and unexpected situation in which the threshold $p_{c,n}$ slightly overshoots the infinite-sub-net value and then decreases as the mesh size increases.   The threshold here is governed by a complicated interplay of connection probabilities for each square, and evidently for intermediate sized systems it is somewhat harder to connect the corners than for larger ones, and this leads to a larger threshold.  In the case of the triangular lattice, where there are fewer connection configurations between the three vertices of one triangle (namely, just $p_0$, $p_2$ and $p_3$), the value of $p_{c,n}$ appears to grow monotonically.

\begin{figure}
\includegraphics[scale = 0.4]{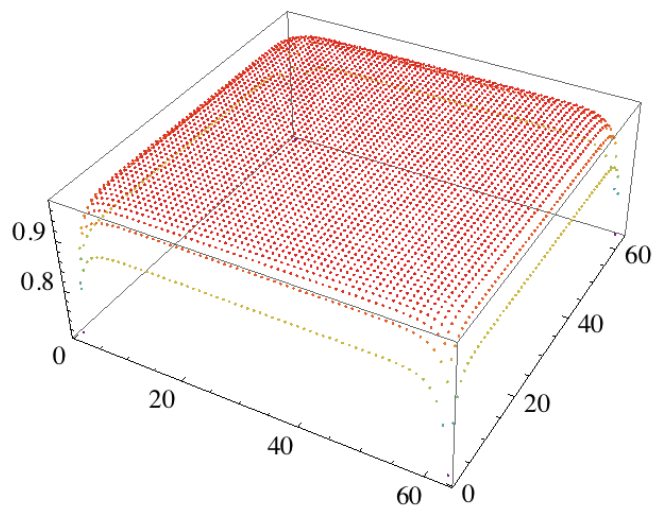}
\caption{\label{hajisquaredensity}  Average density of the infinite (largest) cluster in a $64\times64$ square, at the checkerboard critical point $p_{c,n} = 0.642216$.  At the corners the probability drops to $1/\sqrt{2} = 0.707$ according to Eq.\ (\ref{eq:bottleneck_checkerboard}).}
\end{figure}

To illustrate the general behavior of the systems, we show a typical critical cluster for the $8\times8$ checkerboard system in Fig.\ \ref{Pict8x8squaresBW}.  It can be seen that the checkerboard squares the cluster touches are mostly filled, since the threshold $p_{c,n} = 0.642318$ is so much larger than the square lattice's threshold $p_{c,s} = 0.5$.

\begin{figure}
\includegraphics[scale = 0.4]{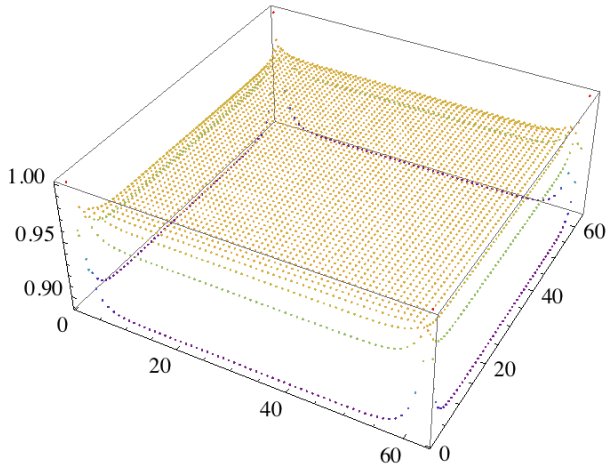}
\caption{\label{haji4density}  Average density of clusters that simultaneously touch all four corners at the checkerboard criticality $p_{c,n} = 0.642216$ for a single $64\times64$ square.}
\end{figure}

In Fig.\ \ref{hajisquaredensity} we show the average density of ``infinite" (large) clusters in a single $64\times64$ square at the checkerboard criticality of $p_{c,n} = 0.642216$, in which case the density drops to $1/\sqrt{2}$ at the corners.  In Fig.\ \ref{haji4density} we show the corresponding densities conditional on the requirement that the cluster simultaneously touches all four corners, so that the density now goes to 1 at the corners and drops to a somewhat lower value in the center because not every site in the system belongs to the spanning cluster.
Similar plots can be made of clusters touching 1, 2, or 3 corners.  At the sub-net critical point $p_{c,s}$, the first two cases can be solved exactly and satisfy a factorization condition \cite{SimmonsKlebanZiff07,SimmonsZiffKleban08}, but this result does not apply at the higher $p_{c,n}$.

The ideas discussed in this paper apply to any system with regular bottlenecks.  Another example is the kagom\'e lattice with the triangles filled with a finer-mesh triangular lattice; this system is studied in Ref.\ \cite{ZiffGu}.

 \section{Acknowledgments}

This work was supported in part by the National Science Foundation Grants No.\ DMS-0553487 (RMZ).  The
authors also acknowledge the contribution of UROP (Undergraduate Reseach Opportunity Program) student 
Hang Gu for his numerical determination of $p_{c,n}$ for the triangular lattice of sub-net size $5\times5$, and thank Christian R. Scullard for helpful discussions concerning this work.

\bibliographystyle{apsrev}%{unsrt}
%\bibliographystyle{unsrt}%{unsrt}
%\bibliography{percolation}% Produces the bibliography via BibTeX.
\bibliography{HajiAkbariZiffv3}
\end{document}